\documentclass[letterpaper,notoc]{CECS}
\title{Asymptotically Anti-de Sitter spacetimes \\ and scalar fields with a
logarithmic branch}
\author{Marc Henneaux$^{1,2}$,  Cristi\'an Mart\'{\i}nez$^{1}$, Ricardo Troncoso$^{1}$ and Jorge Zanelli$^{1}$\footnote{{\it E-mail:} {\tt henneaux@ulb.ac.be, martinez@cecs.cl, ratron@cecs.cl, jz@cecs.cl}}\\ $^{1}$Centro de Estudios Cient\'{\i}ficos (CECS), Casilla 1469, Valdivia, Chile
\\$^{2}$Physique th\'{e}orique et math\'ematique and International
Solvay Institutes, Universit\'{e} Libre de Bruxelles, Campus Plaine C.P.231, B-1050
Bruxelles, Belgium.  }

\preprint{{\tiny CECS-PHY-04/08} \\[-2.5mm] {\tiny ULB-TH-04/15}}

\abstract{
We consider a self-interacting scalar field whose mass saturates the
Breitenlohner-Freedman bound, minimally coupled to Einstein gravity with a
negative cosmological constant in $D\geq 3$ dimensions. It is shown that the
asymptotic behavior of the metric has a slower fall--off than  that  of  pure
 gravity  with  a localized  distribution of  matter, due  to  the  back-reaction  of
the  scalar  field, which  has  a  logarithmic  branch decreasing  as $  r^{-(D-1)/2} \,\ln r$ for large radius  $r$.

We find the asymptotic conditions on the fields which are invariant under
the same symmetry group as pure gravity with negative cosmological constant
(conformal group in $D-1$ dimensions). The generators of the asymptotic
symmetries are finite even when the logarithmic branch is considered but
acquire, however, a contribution from the scalar field.}

\begin{document}
\section{Introduction}

In the presence of matter fields, the asymptotic behavior of the
metric can be different from that arising from pure gravity if the
matter fields do not fall off sufficiently fast at infinity. This
occurs, for instance, in the case of three-dimensional black holes
with electric charge \cite{BTZ,MTZ}, as well as in the presence of
scalar fields \cite{Henneaux:2002wm}. The modified asymptotic form
of the fields can, a priori, modify the original symmetry at
infinity and its associated charges. However, as the example
analyzed in \cite{Henneaux:2002wm} shows, it might be possible to
relax the standard asymptotic conditions without losing the
original symmetry, but modifying the charges in order to take into
account the presence of the matter fields. This effect occurs in a
dramatic way when the mass of the scalar field saturates the
Breitenlohner-Freedman bound \cite{B-F,M-T}, since in this case, a
potentially dangerous logarithmic branch appears in the asymptotic
form of the scalar field.

In this paper we deal with a self-interacting scalar field
minimally coupled to $D$-dimensional Einstein gravity with a
negative cosmological constant, whose B-F saturating mass is
explicitly given by

\begin{equation}
m_{*}^{2}=-\frac{(D-1)^{2}}{4l^{2}}\;,  \label{mBF}
\end{equation}
where $l$ is the AdS radius\footnote{%
Scalar fields with masses lower than this value produce a perturbative
instability of AdS spacetime, as shown in Ref. \cite{B-F} in four
dimensions, and generalized to other dimensions in \cite{M-T}. The case
previously considered in \cite{Henneaux:2002wm} does not saturate the
Breitenlohner-Freedman bound, which is one of the reasons why we focus here
on the case (\ref{mBF}).}.

We prove that the back-reaction produced by the logarithmic branch in the
scalar field requires an asymptotic behavior of the metric which differs
from the standard asymptotic behavior found in \cite
{Henneaux-Teitelboim,Henneaux-D} by the addition of logarithmic terms.
Nevertheless, this relaxed asymptotic metric still preserves the original
asymptotic symmetry, which is $SO(D-1,2)$ for $D>3$, and the conformal group
in two spacetime dimensions for $D=3$ \cite{Brown-Henneaux}. Furthermore,
the conserved charges acquire an extra contribution coming from the scalar
field, and they are finite even when the logarithmic branch is switched on.

We consider the following action
\begin{equation}
I[g,\phi ]=\int d^{D}x\sqrt{-g}\left( \frac{R}{16\pi G}-\frac{1}{2}(\nabla
\phi )^{2}-V(\phi )\right) \;,  \label{action}
\end{equation}
with a potential $V(\phi )$ given by
\begin{equation}
V(\phi )=\frac{\Lambda }{8\pi G}+\frac{m_{*}^{2}}{2}\phi ^{2}+\phi
^{3}U(\phi )\;,  \label{potential}
\end{equation}
where $U(\phi )$\ could be any smooth function\footnote{%
This means that we consider any potential having a negative local maximum at
$\phi =\phi _{0}$, so that $\Lambda =8\pi GV(\phi _{0})$, with $V^{\prime
\prime }(\phi _{0})=m_{*}^{2}$.} around $\phi =0$. Here $G$ is the
gravitational constant, which is $G=(8\pi )^{-1}$ in appropriate units, and
the cosmological constant $\Lambda $ is related to the AdS radius $l$
through $\Lambda =-l^{-2}(D-1)(D-2)/2$.

In order to write down the asymptotic behavior of the fields, the metric is
written as $g_{\mu \nu }=\bar{g}_{\mu \nu }+h_{\mu \nu }$, where $h_{\mu \nu
}$ is the deviation from the AdS metric,
\begin{equation}
d\bar{s}^{2}=-(1+r^{2}/l^{2})dt^{2}+(1+r^{2}/l^{2})^{-1}dr^{2}+r^{2}d\Omega
^{D-2}\;.
\end{equation}

For matter-free gravity, the asymptotic behavior of the metric is given in
\cite{Henneaux-Teitelboim,Henneaux-D,Brown-Henneaux}
\begin{equation}
\begin{array}{lll}
h_{rr} & = & \displaystyle O(r^{-D-1})\;, \\[2mm]
h_{rm} & = & O(r^{-D})\;, \\[1mm]
h_{mn} & = & O(r^{-D+3}).\;
\end{array}
\label{Standard-Asympt}
\end{equation}
Here the indices have been split as $\mu =(r,m)$, where $m$ includes the
time coordinate $t$ plus $D-2$ angles. It is easy to check that the
asymptotic conditions (\ref{Standard-Asympt}) are invariant under $SO(D-1,2)$
for $D>3$, and under the infinite-dimensional conformal group in two
dimensions for $D=3$. The asymptotic behavior of a generic asymptotic
Killing vector field $\xi ^{\mu }$ is given by
\begin{equation}
\begin{array}{lllllll}
\xi ^{r} & = & O(r), &  & \xi _{,r}^{r} & = & O(1) \\
\xi ^{m} & = & O(1), &  & \xi _{,r}^{m} & = & O(r^{-3})
\end{array}
\label{Asympt-Symm}
\end{equation}

The charges that generate the asymptotic symmetries involve only the metric
and its derivatives, and are given by
\begin{equation}
Q_{0}(\xi )=\frac{1}{2}\int d^{D-2}S_{i}\left\{ \bar{G}^{ijkl}(\xi ^{\bot
}g_{kl|j}-\xi _{\;\;|j}^{\bot }h_{kl})+2\xi ^{j}\pi _{j}^{\;\;i}\right\} \;,
\label{Q0}
\end{equation}
where $G^{ijkl}=\frac{1}{2}g^{1/2}(g^{ik}g^{jl}+g^{il}g^{jk}-2g^{ij}g^{kl})$%
, and the vertical bar denotes covariant differentiation with respect to the
spatial AdS background. From (\ref{Standard-Asympt}) it follows that the
momenta possess the following fall-off at infinity
\begin{equation}
\pi ^{rr}=O(r^{-1}),\quad \pi ^{rm}=O(r^{-2}),\quad \pi ^{mn}=O(r^{-5})\;,
\end{equation}
and hence, the surface integral (\ref{Q0}) is finite.

The Poisson brackets algebra of the charges yields the AdS group for $D>3$
and two copies of the Virasoro algebra with a central charge given by
\begin{equation}
c=\frac{3l}{2G}  \label{Central Charge}
\end{equation}
in three dimensions \cite{Brown-Henneaux}

\section{Switching on the scalar field}

The asymptotic conditions (\ref{Standard-Asympt}) hold not just in the
absence of matter but also for localized matter fields which fall off
sufficiently fast at infinity, so as to give no contributions to the surface
integrals defining the generators of the asymptotic symmetries. The scalar
field would not contribute to the charges if it goes as $\phi \sim
r^{-((D-1)/2+\varepsilon )}$ for large $r$. However, when the mass of the
scalar field is given by (\ref{mBF}), saturating the Breitenlohner-Freedman
bound, the fall-off of the field is slower and generally produces a strong
back-reaction that relaxes the asymptotic behavior of the metric.

In this case, the leading terms for $h_{\mu \nu }$ and $\phi $ as $%
r\rightarrow \infty $ read

\begin{eqnarray}
&&
\begin{array}{lll}
\phi  & = & \!\displaystyle \frac{1}{r^{(D-1)/2}}\left( a+b\ln \left(
r/r_{0}\right) \right) +O\left( \frac{\ln (r/r_{0})}{r^{(D+1)/2}}\right)  \\%
[3mm]
&  &
\end{array}
\nonumber \\
&&
\begin{array}{lll}
h_{rr} & = & \!\displaystyle -\frac{(D-1)l^{2}b^{2}}{2(D-2)}\;\frac{\ln
^{2}\left( r/r_{0}\right) }{r^{(D+1)}} \\[3mm]
& + & \!\displaystyle \frac{l^{2}(b^{2}-(D-1)ab)}{D-2}\;\frac{\ln \left(
r/r_{0}\right) }{r^{(D+1)}}+O\left( \frac{1}{r^{(D+1)}}\right)  \\[3mm]
h_{mn} & = & \!\displaystyle O\left( \frac{1}{r^{(D-3)}}\right)  \\[3mm]
h_{mr} & = & \!\displaystyle O\left( \frac{1}{r^{(D-2)}}\right)  \\[3mm]
&  &
\end{array}
\end{eqnarray}
where $a=a(x^{m})$, $b=b(x^{m})$, and $r_{0}$ is an arbitrary constant.

Indeed, for an asymptotically AdS spacetime, the leading terms in
the metric determine the fall-off of $\mathbf{\phi }$ through its
field equation, and in turn, a back-reaction with logarithmic terms in $%
h_{rr}$ is obtained by solving the constraints.

When the logarithmic branch of the scalar field is switched on ($b\neq 0$),
these relaxed asymptotic conditions still preserve the original asymptotic
symmetry, provided
\[
a=-\frac{2}{(D-1)}\, b \, \ln (b/b_{0})\;,
\]
where $b_{0}$ is a constant. Not that for $a=0$, $b$ must be a constant. For
$b=0$, the asymptotic symmetry does not impose restrictions on $a$.

Using the Regge-Teitelboim approach \cite{Regge-Teitelboim} the
contributions of gravity and the scalar field to the conserved charges, $%
Q_{G}(\xi )$ and $Q_{\phi }(\xi )$ are given by
\begin{eqnarray}
\delta Q_{G}(\xi ) &=&\frac{1}{2}\int d^{D-2}S_{l}\left[ G^{ijkl}(\xi ^{\bot
}\delta g_{ij;k}-\xi _{,k}^{\bot }\delta g_{ij})\right.  \nonumber \\
&+&\left. \frac{1}{2}\int d^{D-2}S_{l}(2\xi _{k}\delta \pi _{kl}+(2\xi
^{k}\pi ^{jl}-\xi ^{l}\pi ^{jk})\delta g_{jk})\right]  \label{DeltaQGgen} \\
\delta Q_{\phi }(\xi ) &=&-\int d^{D-2}S_{l}\left( \xi ^{\bot
}g^{1/2}g^{lj}\partial _{j}\phi \delta \phi +\xi ^{l}\pi _{\phi }\delta \phi
\right) \;.  \nonumber  \label{DeltaQPhigen}
\end{eqnarray}
Making use of the the relaxed asymptotic conditions, the momenta at infinity
are found to be
\begin{eqnarray}
\pi ^{rr}=O(r^{-1}),\quad &\pi ^{rm}&=O(r^{-2}),\quad \pi
^{mn}=O(r^{-5}\ln^{2}(r)), \\
&\pi _{\phi }&=O(r^{(D-7)/2}\ln(r))\,,
\end{eqnarray}
and hence Eqs. (\ref{DeltaQGgen}), and (\ref{DeltaQPhigen}) acquire the form
\begin{eqnarray}
\delta Q_{G}(\xi ) &=&\left. \delta Q_{0}(\xi )\right| _{\phi =0} +\frac{1}{2%
}\int d\Omega ^{D-2}\frac{\xi ^{t}}{l^{2}}\left[ -\frac{D-1}{2}\delta
b^{2}\ln^{2}(r/r_{0}) \right.  \nonumber \\
&+& \displaystyle \left. \delta (b^{2}-(D-1)a b)\ln(r/r_{0})\right]
\label{DeltaQGvar} \\
\delta Q_{\phi }(\xi ) &=&-\frac{1}{2}\int d\Omega ^{D-2}\frac{\xi ^{t}}{%
l^{2}}\left[ (2b-(D-1)a)\delta a\right.  \nonumber \\
&&\left. -\frac{D-1}{2}\delta b^{2}\ln^{2}(r/r_{0})+\delta (b^{2}-(D-1)a
b)\ln(r/r_{0})\right] \;.  \label{DeltaQPhivar}
\end{eqnarray}

Asymptotically, $\xi ^{t}\sim O(1)$ and therefore $\delta Q_{G}$, and $%
\delta Q_{\phi }$, both pick up logarithmic divergences, but these divergent
pieces exactly cancel out. Thus, the total variation is well defined and the
total charge $Q=Q_{G}+Q_{\phi }$ can be integrated, obtaining
\begin{equation}
Q(\xi )=Q_{0}(\xi )+\int d\Omega ^{D-2}g^{1/2}\frac{r}{l^{2}}\xi ^{\bot
}\left\{ \frac{D-1}{8}\phi ^{2}+\frac{r^{2}}{2(D-1)}(\partial _{r}\phi
)^{2}\right\} \,,  \label{Qtotal}
\end{equation}
with $Q_{0}(\xi )$ given by (\ref{Q0}). Consequently, the conserved charges
acquire an extra contribution coming from the scalar field, and they are
finite even when the logarithmic branch is switched on. Note that in the
case $b=0$, the asymptotic behavior of the metric reduces to the standard
one (\ref{Standard-Asympt}), and the original asymptotic symmetry is
preserved, but nevertheless, the charges (\ref{Qtotal}) still give a
non-trivial contribution coming from the scalar field. One should expect
that a similar expression for $Q(\xi )$ in Eq.(\ref{Qtotal}) could also be
found using covariant methods as in \cite{Glenn}

The algebra of the charges (\ref{Qtotal}) is identical to the standard one,
namely AdS for $D>3$ \cite{Henneaux-Teitelboim,Henneaux-D}, and two copies
of the Virasoro algebra with the same central extension for $D=3$ \cite
{Brown-Henneaux}. This can be readily obtained following Ref \cite
{Brown-Henneaux2}, where it is shown that the bracket of two charges
provides a realization of the asymptotic symmetry algebra with a possible
central extension. The central charge can be determined by the variation of
the charges in the vacuum, represented here by AdS spacetime with $\phi =0$.

\section{Discussion}

The presence of matter fields with nontrivial asymptotic behavior
has generically two effects: It gives rise to a back reaction that
modifies the asymptotic form of the geometry, and it generates
additional contributions to the charges that depend explicitly on
the matter fields at infinity which are not already present in the
gravitational part. These two effects were observed in 2+1
dimensions in Ref. \cite{Henneaux:2002wm}, and it is also seen
here in the presence of a scalar field with a logarithmic branch.
Furthermore, even when the logarithmic branch is switched off
($b=0$ in Eq.(\ref{DeltaQPhivar})), the scalar field still gives a
contribution to the charge, even though the asymptotic form of the
metric is unchanged.

As shown here, the presence of the logarithmic branch in the scalar field is
consistent with asymptotically AdS symmetry (both kinematically in the sense
that the boundary conditions are preserved, and dynamically in the sense
that the associated charges are finite), provided one takes into account the
back reaction in the metric. In other words, it would be inconsistent to
treat the scalar field as a probe in this case. It should also be stressed
that it is only the sum of the gravitational contribution \textit{and} of
the scalar field contribution which is conserved and which defines a
meaningful AdS charge. Each term separately may vary as one makes asymptotic
AdS time translations.

We note that the AdS charges of metric-scalar field configurations
with a logarithmic branch can also be computed through the method
of holographic renormalization, as explicitly performed in
\cite{BFS1,BFS2} for $D=5$.  Those papers contain a detailed
discussion of the AdS/CFT correspondence in this context. The
counterterms needed to render the action finite explicitly depend
on the scalar field.  This was also shown for the $D=3$ case of
\cite{Henneaux:2002wm} in the article \cite{Gegenberg:2003jr}.

Another question that deserves further study is the positivity of the AdS
energy in the wider context considered here (positivity has been proved
under standard, supersymmetric boundary conditions in \cite
{Abbott:1981ff,Gibbons:aq} following the asymptotic flat space derivation of
\cite{Deser:hu,Witten:mf}). A related issue is whether the slower fall-off
of the metric and the logarithmic tail of the scalar field are compatible
with supersymmetry. In that respect, the spin $1/2$-partner of the scalar
field should play a crucial role and is expected to contribute to the
supercharge. This question is under current investigation.

Scalar fields are involved in a recent controversy concerning
cosmic censorship \cite{Controversy}. In particular, in Ref.
\cite{Hertog:2003xg}, the cosmic censorship conjecture is
challenged through the evolution of a scalar field in five
dimensions, whose logarithmic branch is switched on. One may
consider the following set of initial conditions without
introducing a cut-off: $\phi =\phi _{0}$, for $r<R_{0}$, and $\phi
=b_{0}r^{-2}\ln r$, for $r>R_{0}$, so that $\phi
_{0}=b_{0}R_{0}^{-2}\ln R_{0}$. The mass of this configuration is
obtained through Eq. (\ref{Qtotal}), with $\xi =\partial _{t}$,
and reads
\begin{equation}
M=\frac{\pi ^{2}b_{0}^{2}}{4l^{2}}(1+4\ln ^{2}R_{0})\;,  \label{MHHM}
\end{equation}
which is manifestly finite and positive. It is argued in \cite
{Hertog:2003xg}, that a naked singularity will develop at
$r=R_{s}\sim \phi _{0}^{2/3}R_{0}$, for $\phi _{0}<<1$. However,
the mass as given by (\ref{MHHM}) has an associated Schwarzschild
radius $r_{+}$  satisfying $r_{+}>>R_{s}$, so that it encloses the
singularity. This may help resolve the controversy.

\acknowledgments

We thank T. Hertog, V. Hubeny, and M. Rangamani for useful discussions and
enlightening comments. This research is partially funded by FONDECYT grants
1020629, 1010446, 1010449, 1010450, 1040921, 7010446, and 7010450. The work of MH is
partially supported by IISN - Belgium (convention 4.4505.86), by a ``P\^{o}%
le d'Attraction Universitaire'' and by the European Commission RTN programme
HPRN-CT-00131, in which he is associated to K. U. Leuven. C. M. and R. T. wish to thank
the kind hospitality at the Physique th\'{e}orique et math\'ematique and International
Solvay Institutes, Universit\'{e} Libre de Bruxelles. The generous
support to Centro de Estudios Cientificos (CECS) by Empresas CMPC is also
acknowledged. CECS is a Millennium Science Institute and is funded in part
by grants from Fundaci\'{o}n Andes and the Tinker Foundation.

\end{document}